\documentclass{article}
\usepackage{enumitem}
\usepackage{hyperref}
\usepackage{setspace}
\usepackage{graphicx}
\usepackage{caption}
\usepackage{amsmath,amsthm}
\usepackage[left=3cm,right=4cm]{geometry}
\renewcommand{\qed}{\hfill \mbox{\raggedright \rule{0.1in}{0.1in}}}
\date{}
\begin{document}
\title{\bf  Comments on a recently proposed Privacy Preserving Lightweight Biometric Authentication System for IoT Security }
\author{SrinivasaRao SubramanyaRao and Enrique Argones Rua}
\maketitle
\begin{center}
	\textit{(This paper has been submitted to IEEE for possible publication)}
\end{center}
\begin{abstract}
In this paper, we show that a recently published lightweight adaptation of a Fingerprint matching algorithm called the \textit{Minutia Cylinder-Code} may not be secure as intruders may be able to illegitimately yet successfully authenticate themselves to the system under consideration. We also show that the lightweight adaptation has other privacy related vulnerabilities that make it unsuitable for use in Biometrics. We make it clear that we are neither investigating nor commenting on the security of the original Minutia Cylinder-Code algorithm by itself, rather we highlight the vulnerabilities of the lightweight adaptation. In the process of doing this, we provide a high-level overview of the role of one-way functions in cryptography and biometrics to provide a context to the aforementioned lightweight algorithm and its deficiencies.\\
\end{abstract}
\section{Introduction}
\noindent In \cite{Cappelli}, Cappelli, Ferrara and Maltoni proposed a novel Fingerprint recognition technique called the \textit{Minutia Cylinder-Code} (MCC) and is based on $3D$ data structures called Cylinders. In an article in the March 2019 issue of IEEE Communication Magazine \cite{WYang}, Yang, Wang, Zheng, Yang and Valli  presented a new Privacy Preserving Lightweight Biometric system for resource constrained IoT devices, where an improvement of the aforementioned MCC  algorithm was proposed. As it is not our intention to investigate the original MCC algorithm in this paper, we refer the readers to \cite{Cappelli} for details on the MCC technique.\\

\noindent In this article, we focus on the privacy preserving lightweight adaptation of the MCC algorithm proposed in \cite{WYang} where the authors intend to improve the \textit{energy efficiency} of biometric systems in an IoT environment and call it the \textit{green issue}. They list various Biometric recognition/authentication algorithms in the literature and then select the MCC algorithm before proceeding to propose improvements to this algorithm with an intention to enhance the security and privacy aspects of the system while at the same time make it lightweight and thus energy efficient. They call their main contribution as \textit{A Block Logic Operation Based Lightweight Biometric System} which we abbreviate as BLOBLBS in this article for convenience. The authors focus on the authentication aspects of the Biometric system under consideration and they write in their paper \cite{WYang}\\

\indent \textit{      } \textit{\dots If an imposter rather than the actual user tries to access the device, the authentication mechanism should be able to detect it and prevent such intrusion\dots}\\

\noindent While we focus on the aforementioned BLOBLBS algorithm in this paper, we refer the reader to \cite{WYang} for other details such as experimental results and analysis. Before we focus on the BLOBLBS algorithm, we take a detour and provide with a brief overview of \textit{one-way functions} and its utility in security areas such as cryptography and biometrics. This detour is intended to provide a broader context in which the BLOBLBS algorithm was designed (The BLOBLBS algorithm is itself an example of a one-way function as we will see later in this paper).  The rest of this paper is structured as follows: After providing an overview of \textit{one-way functions} in Section \MakeUppercase{\romannumeral 2}, we outline the BLOBLBS algorithm in Section \MakeUppercase{\romannumeral 3} and then proceed to show in Section \MakeUppercase{\romannumeral 4} that the authentication mechanism based on this algorithm can be totally compromised (compromised in 100\% of the cases). We also show in Section \MakeUppercase{\romannumeral 4} that other vulnerabilities make the BLOBLBS algorithm unsuitable for use in Biometrics. We conclude in Section \MakeUppercase{\romannumeral 4} with a timely remainder of the so called \textit{Schneier's Law}.
\section{One way Functions}
\noindent One way functions, as the name indicates, is a mathematical function $y=f(x)$ such that it is easy to compute $y$ given the input $x$, but not easy to do the reverse (that is, given $y$, it is not easy to determine the corresponding value of $x$).\\

\noindent One way functions are important in some areas of computer security, for example in \textit{Cryptography}. Cryptography enables people to communicate using a public communication channel such that an adversary may learn very little on what is being communicated even though the public communication channel, by default is insecure. Cryptography also enables authentication and message integrity. Cryptography is heavily reliant on one-way functions to achieve its objectives. An instance of a one-way function useful in cryptography is as follows: While one can \textit{quickly} or \textit{efficiently} multiply two prime numbers $p$ and $q$ (a prime number is a positive integer which is not divisible by any other integer other than 1 and itself) to get the product $n=p*q$, it takes considerably more time to discover the two primes $p$ and $q$ given just $n$ and at the same time both $p$ and $q$ are unknown. This is known as the problem of \textit{factorisation} and as the size of the primes becomes larger, it becomes a \textit{computationally difficult} problem to discover $p$ and $q$. The notions of \textit{quickly/efficiently} and \textit{computationally difficult} are to be understood in the context of \textit{computational complexity theory} and a formal and rigorous treatment of these notions can be readily found in the literature in this area \cite{MSipser, JEHopcroft, MRGarey}. The computational difficulty should hold in most cases and not confined to worst-case test scenarios. Off course, legal users of the system should not be confronted with this computational difficulty whereas the adversary should, resulting in a computational gap between legal users and adversaries. The well-known \textit{RSA} cryptosystem depends on the computational difficulty of factorisation. Other such computationally difficult problems that are useful in cryptography include the so called \textit{Discrete Logarithm Problem (DLP)} that is at the core of encryption schemes such as the ElGamal scheme and digital signatures schemes such as the widely employed \textit{Digital Signature Algorithm (DSA)}. The DLP, while very informally is the task of reversing the exponentiation operation in a finite set with an intention to find the exponent, requires more mathematical machinery to describe clearly and rigorously than that required for the factorization problem. As it is not our intention to cover the requisite mathematical machinery in this paper, we refer the reader to any of the excellent books on cryptography, for example \cite{DStinson, Goldreich}, to explore the details. Chapter-2 in \cite{Goldreich} contains a rigorous analysis of one-way functions. The \textit{elliptic curve} (informally, a cubic curve satisfying some properties) analog of the DSA is used in the now famous Bitcoin cryptocurrency and a few other such \textit{Blockchain} based applications. It has to be mentioned here that no one knows if one way functions actually exist!. One-way functions such as factorisation and the DLP are only conjectured to be one-way and the one-wayness of these functions have been the subject of intense research by mathematicians. As mathematicians spend more and more effort trying to efficiently reverse these one-way functions without success, the confidence in the assumption that it is indeed difficult to reverse these one-way functions increases. Thus, one would expect more confidence in the alleged one-wayness of the factorisation problem than that of the so called \textit{isogeny problem} as the isogeny problem is a new entrant to this category of problems whereas factorisation is a much older problem. (that is, mathematicians have spent more time trying to solve the factorisation problem efficiently than the isogeny problem). The isogeny problem is constructed using elliptic curves but this is not the same as that of the elliptic curve analog of the DLP problem referred to above. Success for any mathematician in cracking any one of these afore mentioned problems will certainly make her/him famous, but will also be disastrous for cryptographic security and every business depending on cryptographic security. Luckily for the security community, this has not happened thus far in the case of factorisation and the DLP. Nevertheless, if an intruder is able to somehow efficiently factorise, then she/he can certainly compromise the RSA cryptosystem, but so far there are no known efficient algorithms for factorisation. On the contrary, no one knows if there are ways to compromise RSA without being able to efficiently factorise. \textit{Cryptographic hash functions}, other than the ones that have been broken thus far such as MD5 and SHA1, are another example of one-way functions. It is now known that alleged one way functions such as factorisation and the DLP (including the elliptic curve analog) are vulnerable to quantum computer attacks whereas the new entrant, the Isogeny problem is thought to be resistant to these attacks. The anticipation of Quantum computers materializing in the near future is the reason for \textit{Post-Quantum Cryptography} \cite{Bernstein} being the subject of intense research currently and this area includes the aforementioned isogeny based cryptography and other areas such as \textit{Lattice-based cryptography} and \textit{code-based cryptography}.\\

\noindent In Biometrics (the metrics of human characteristics such as a fingerprint, face, eyes etc used to identify people) too, a so called \textit{non-invertible} function is utilized to create a distorted or a transformed version of a Biometric and this distorted version is then stored in the biometric database during the process of registering a new user to the system, similar to the process of registering a new user in a password based authentication system. It is not uncommon in the Biometric literature to call this non-invertible function as a one-way function \cite{VMPatel}. The reason for performing this transformation is that if the original biometric template is stolen, then the owner of that biometric cannot use it ever again.  After registration and during the process of authenticating oneself, the user's biometric is translated using the same non-invertible/one-way function that was used during registration and this is then compared with the transformed version stored in the database. Just as in the case of passwords, where the plaintext password by itself is never stored anywhere, the original Biometric is never stored anywhere and thus the danger of loosing the original Biometric template is minimized. This enables biometrics to be revoked and reissued just as a current password can be \textit{canceled} and a new password can be setup when the current password has been compromised. This topic is thus studied under the umbrella of a research area called \textit{Cancelable Biometrics} which signifies that the transformed Biometrics can be canceled when required and a new transformed version reissued. Cancelable Biometrics is one of the ways to protect original Biometric templates. A good overview of template protection techniques can be found in \cite{RaneS}.\\
 
\noindent Even though the term \textit{one-way function} is used both in the cryptography and cancelable biometrics literature, they don't mean exactly the same across the two areas. In cryptography, when a successful attempt is made to reverse the one-way function, the solution has to exactly match the initial inputs to the one-way function, whereas in biometrics the one-way function is usually a \textit{many-to-one} mapping and the hope is that the original biometric, that is intended to be hidden, is just one of many possible inputs to the one-way function. Thus even if the one-way function is compromised by someone who attempts to reverse the one-way function, it is hoped that the original input (which is the original Biometric) to the one-way function is not recovered by the adversary thereby minimizing the chances of the original biometric being stolen. However, if the one-way function is reversed resulting in an exact or even just a close match with the original biometric, then the authentication mechanism utilizing this one-way function can be compromised and an intruder can illegally but successfully authenticate herself/himself to the biometric system. If the intruder is able to reverse the one-way function and obtain the original input, then the original biometric is revealed to the intruder. If the intruder is able to reverse the one-way function but not obtain the original input, then the original biometric is hidden from the intruder, but can still authenticate herself/himself to the system. Conceptually, this is similar to a password based authentication system where the hash of the password is stored in the database during registration and during subsequent authentication attempts, the user supplies her/his password and then the hash of the supplied password is compared to that stored in the database and if an exact match occurs, then the user is positively authenticated. Even if the password supplied by an intruder is different from the original one used by the legitimate user, if the hash function produces the same output as it did for the legitimate user (that is, there is a hash-function collision), then the intruder will obtain access to the system, though illegally, without even knowing the original password. To emphasize this point in the context of biometrics, we quote Nandakumar and Jain from a recent paper \cite{KNanda}. They write\\

\indent \dots \textit{it may not be necessary to exactly recover the original template from the protected biometric reference. Instead, it is sufficient for the attacker to obtain a close approximation (also known as a pre-image), which can be replayed to the system to gain illegitimate access\dots}\\

\noindent There are additional important properties that a one-way function in cancelable biometrics should satisfy. Transformations introduced by the same one-way function on original biometrics of the same individual authenticating to the same entity should be close to each other such that the authentication system can successfully match the different transformations as belonging to the same individual and positively authenticate that individual. However two or more transformations introduced by the one-way function on original biometrics of the same individual authenticating to two or more different entities should not match with each other, else this can result in cross-matching and can lead to an individual's privacy being compromised. Moreover, transformations produced by the one-way function on original biometrics of different individuals should not match with each other. For more details of properties of one-way functions utilized in cancelable biometrics, we refer the reader to \cite{SChikkerur, AKJain}. It is these additional properties that one-way functions should satisfy that make the nature of one-way/non-invertible functions in cancelable biometrics different to the one-way functions used in cryptography. Thus a cryptographic hash function which is a one-way function cannot be directly employed in cancelable biometrics, as even small variations in input can lead to large differences in the output.\\

\noindent While there are well understood techniques from computational complexity theory to measure the difficulty of reversing one-way functions and is considered standard in cryptography, these methods do not suffice in biometrics and moreover there are no universally accepted techniques to quantify template protection algorithms \cite{KNanda} in Biometrics. Nagar and Jain introduced a measure of non-invertibility in the context of cancelable biometrics in \cite{ANagar}. Metrics for benchmarking biometric template protection algorithms were outlined in \cite{KSimoens}.  Further efforts including standardization in this direction by the biometrics community are summarized by Rane in \cite{SRane}. Yet in addition to the metrics referred to above, the strength of non-invertibility depends on the computational complexity of the best available algorithms to reverse the one-way function \cite{KNanda}, just as in cryptography.\\

\noindent Commonly known one-way functions in the cancelable biometrics literature include \textit{geometric transforms, random projections} and \textit{biohashing}. A class of one-way functions suitable for Biometrics are called \textit{continous one-way} functions by Grigoriev and Nikolenko in \cite{DGrigoriev}. They write\\

\indent \textit{\dots We understand continuity in the regular mathematical sense: a continuous function maps close points of a Euclidean space to close points of another Euclidean space. This setting makes perfect sense for Biometric applications \dots}\\

\noindent Just as in the case of the DLP in the cryptography domain, description of one-way functions useful in biometrics referred to above requires more mathematical machinery than we intend to develop in this paper. We refer the reader to \cite{VMPatel} for more details and a summary of these one-way functions.\\

\noindent In the light of the above discussions, it would not be an unreasonable hypothesis to state that constructing good one way functions for cancelable biometrics may be more involved that those suitable for cryptography. \\

\noindent Now the algorithm that we referred to in Section \MakeUppercase{\romannumeral 1} of this paper (i.e., the BLOBLBS algorithm proposed in \cite{WYang}) is an instance of a one-way function in cancelable biometrics and this algorithm is the subject of focus in subsequent sections in this paper.\\

\section{The BLOBLBS Algorithm}
\noindent The BLOBLBS algorithm was designed to enhance security pertaining to user authentication. In fact, the authors in \cite{WYang} write\\

\indent \textit{\dots The proposed lightweight biometric system aims to offer user authentication to IoT devices with enhanced security \dots}\\

\noindent The BLOBLBS as proposed in \cite{WYang} can be outlined as follows:\\

\noindent In the biometric enrollment stage, the minutiae is first extracted and then a binary vector is constructed from this extraction. For simplicity, let us assume that this binary vector is saved in a file $F_{1}$. Further the authors define a transformation called the \textit{Block Logic Operation}(BLO) which transforms $F_{1}$ into another binary vector that can be saved in another file $F_{2}$ and this consists of the transformed template. The transformation is such that the size of $F_{1}$ is greater than $F_{2}$. Biometric authentication consists of two stages - \textit{enrollment} and \textit{matching}. In the matching stage, the three steps of minutiae extraction, binary feature generation and application of the BLOBLBS algorithm are performed exactly as in the enrollment stage and then the corresponding transformed templates ($F_{2}$) from the two stages are compared to determine if there is a match.\\

\noindent The transformation from $F_{1}$ to $F_{2}$ is as follows: The file $F_{1}$ is sliced into many segments with each segment containing, a fixed odd number, say 5 bits $(b_{1},b_{2},b_{3},b_{4}$ and $b_{5})$ each. Each such segment can be written in the form of a 5-tuple $(b_{1},b_{2},b_{3},b_{4},b_{5})$. Each such 5-tuple slice in the file $F_{1}$ is transformed to a 4-tuple as $(b_{1} \oplus b_{3},b_{2} \oplus b_{3},b_{4}\oplus b_{3},b_{5}\oplus b_{3}) = (c_{1},c_{2},c_{3},c_{4})$ where $\oplus$ denotes the XOR operation and the sequence of 4-tuples concatenated corresponding to the sequence of 5-tuples in $F_{1}$ make up the file $F_{2}$.\\

\noindent The authors in [2] describe a file $F_{1}$ of size $1792$ bits. Without loss of generality and for convenience, we take this size to be $1795$ bits. If $F_{1}$ is of size 1795 bits, since there are 359 slices of 5 bits each in $F_{1}$, the size of $F_{2}$ will be $1795 - 359 = 1436$ bits. Thus, in general, the size of the biometric feature vector is reduced and the authors in \cite{WYang} write \\

\indent \textit{      } \textit{\dots the key is choosen from the biometric feature itself and discarded after use, so that the risk caused by key loss can be avoided \dots}\\

\noindent However, the question to be asked is, does the proposed algorithm enhance the security of user authentication as claimed in \cite{WYang} and is it suitable for cancelable biometrics? In the next section we show that the authentication mechanism, when the BLOBLBS algorithm is employed, can be effortlessly compromised. In other words, instead of enhancing the security of the authentication mechanism, the BLOBLBS algorithm reduces the security to such an extent that an imposter does not encounter any resistance when trying to impersonate herself/himself. In addition, we provide reasons to show that the algorithm does not lend itself well to the requirements of one-way functions in cancelable biometrics.  
\section{Issues with the BLOBLBS Algorithm and its Vulnerabilities}
\noindent To perform a rigorous security analysis of a Biometric authentication system (or for that matter any other security system), the system needs to be adequately described.\\

\noindent The need for a rigorous treatment of cryptography is well explored in Chapter-1 of \cite{Goldreich}. Goldreich writes in \cite{Goldreich}\\

\textit{\dots it is implicitly assumed that the basic concepts of cryptography (e.g., secure encryption) are self-evident (because they are so natural) and that there is no need to present adequate definitions. The fallacy of that assumption is demonstrated by the abandon of papers (not to mention private discussions) that derive and/or jump to wrong conclusions concerning security. In most cases these wrong conclusions can be traced back to implicit misconceptions regarding security that could not have escaped the eyes of the authors if they had been made explicit.. \dots}\\

\noindent and backs the above argument with a well-known example.  Rigorous treatment in the area of Biometric security is useful too, as evidenced in a 2005 paper \cite{MJAtallah} where Atallah et al present a lightweight scheme for biometric authentication and in the process of doing so, they define the adversary model stating clearly what the adversary can and cannot do, paving way for a rigorous security analysis of the authentication protocols presented further in their paper. Similarly, in the case of BLOBLBS, an adequate definition of the adversary strength and its goals could enable a good evaluation of system security enabled by the algorithm.\\
 
\noindent Nevertheless, in the absence of a clearly defined adversary model accompanying the  BLOBLBS algorithm in \cite{WYang} and with the information available to us in that paper, we proceed to perform a security analysis of this algorithm and show that there are security and privacy issues with the algorithm. \\

\noindent We wish to state here that even if the adversary model were to be precisely defined, we anticipate that the BLOBLBS algorithm would still be vulnerable. The security and privacy of BLOBLBS depends on the answers to the following questions:\\
\indent \textit{(i) Given a segment in $F_{2}$, how difficult or easy is it for the adversary to reconstruct a corresponding segment in $F_{1}$.}? \\
\indent \textit{(ii) Can there be a cross match of the transformed templates between various databases (in this case, various IoT devices) thus enabling identification of the same individual enrolled across databases compromising privacy? }\\
\indent \textit{(iii) Just as in the case of passwords, is it easy to revoke/cancel a template and issue a new protected template in its place based on the same feature?}\\
\indent \textit{(iv) Is recognition accuracy impacted in the transformed domain?}\\

\noindent We answer these questions below:\\

\noindent (i) Given $F_{2}$, it would suffice if we can efficiently construct a segment in $F_{1}$ that maps to the corresponding one in $F_{2}$. This is easy to do. Given a 4-tuple $(c_{1},c_{2},c_{3},c_{4})$ in $F_{2}$, one can construct the corresponding 5-tuple in $F_{1}$. In fact, two such 5-tuples in $F_{1}$ map to the same 4-tuple $(c_{1},c_{2},c_{3},c_{4})$ in $F_{2}$. These two 5-tuples are $(c_{1} \oplus 0,c_{2} \oplus 0,0, c_{3} \oplus 0,c_{4}\oplus 0)$ and $(c_{1} \oplus 1,c_{2} \oplus 1,1, c_{3} \oplus 1,c_{4}\oplus 1)$.\\

\noindent Given that the notation $\overline{A}$ denotes the complement of the boolean variable $A$, it is easy to see that $\overline{c_{i} \oplus 0} = c_{i} \oplus 1$ and $\overline{c_{i} \oplus 1} = c_{i} \oplus 0$ for $1 \le i \le 4$.  This is because, in general for boolean variables $A$ and $B$, it is easy to verify that $\overline{A \oplus B} = A \oplus \overline{B}$. In other words, each of the elements in one of the 5-tuples above is a complement of the corresponding element in the other 5-tuple.\\ 

\noindent Given that either of the two 5-tuples above $(c_{1} \oplus 0,c_{2} \oplus 0,0, c_{3} \oplus 0,c_{4}\oplus 0)$ or $(c_{1} \oplus 1,c_{2} \oplus 1,1, c_{3} \oplus 1,c_{4}\oplus 1)$, when subject to the BLO transformation, can produce the 4-tuple $(c_{1},c_{2},c_{3},c_{4})$ and since the 4-tuple  $(c_{1},c_{2},c_{3},c_{4})$ can assume 16 possible values, it is easy to list the corresponding two 5-tuples for each of the 16 possible 4-tuples. For example, for the 4-tuple $(1,0,1,0)$, one can construct the corresponding two 5-tuples by computing $(1 \oplus 0, 0 \oplus 0, 0, 1 \oplus 0, 0 \oplus 0 ) = (1,0,0,1,0)$ and $(1 \oplus 1, 0 \oplus 1, 1, 1 \oplus 1, 0 \oplus 1 ) = (0,1,1,0,1)$. Thus, one can construct a $F_{1}$ segment corresponding to a $F_{2}$ segment and the security of the system can thus be compromised.\\

\indent \textit{ }\\
\begin{tabular}{ |c | c | c| c| }
	\hline 
	$F_{2}$ (4 bits) & $F_{1}$ (5 bits) &$F_{2}$ (4 bits) & $F_{1}$ (5 bits)  \\  \hline
	$0000$ & $00000$  &$0001$ & $00001$  \\ 
	& $11111$  &       & $11110$  \\ \hline       
	$0010$ & $00010$  &$0011$ & $00011$  \\ 
	& $11101$  &       & $11100$  \\ \hline       
	$0100$ & $01000$  &$0101$ & $01001$  \\ 
	& $10111$  &       & $10110$  \\ \hline
	$0110$ & $01010$  & $0111$ & $01011$ \\ 
	& $10101$  &        & $10100$  \\ \hline
	$1000$ & $10000$  & $1001$ & $10001$  \\ 
	& $01111$  &        & $01110$  \\ \hline
	$1010$ & $10010$  & $1011$ & $10011$  \\ 
	& $01101$  &        & $01100$  \\ \hline
	$1100$ & $11000$  & $1101$ & $11001$  \\ 
	& $00111$  &        & $00110$  \\ \hline
	$1110$ & $11010$  & $1111$ & $11011$  \\ 
	& $00101$  &        & $00100$  \\ \hline 
\end{tabular}\\
\indent \textit{ }\\
\indent \textit{ }\\
\noindent For completeness, we provide the mapping for each of the 16 possible 4-tuples, in the above table.\\

\noindent \textbf{Example}:\\
(a) Let $F_{2}$ = $1010$. Thus $F_{2}$ has $n=1$ block of $4$ bits and from the above table, $F_{1}$=$10010$ or $01101$. Thus, there are $2$ possibilities for $F_{1}$.\\
\noindent (b) Let $F_{2}$ = $10101000$. Thus $F_{2}$ has $n=2$ blocks of 4 bits each. The two blocks are $1010$ and $1000$.  For the block $1010$ in $F_{2}$, the possibilities for the corresponding block in $F_{1}$ are $10010$ and $01101$. Similarly for the block $1000$ in $F_{2}$ = $1000$, the possibilities for the corresponding block in $F_{1}$ are $10000$ and $01111$.\\
Thus when $F_{2}$ = $10101000$,\\ $F_{1}$= $10010 10000$ or $10010 01111$ or $01101 10000$ or $01101 01111$. \\ 
\noindent Thus there are $4$ possibilities for $F_{1}$ as each block in $F_{1}$ has two possibilities and there are $n=2$ blocks. Any of the 4 possibilities above for $F_{1}$ will produce $F_{2}$=10101000. \qed \\

\noindent In general, if there are $n$ blocks in $F_{1}$ (or $F_{2}$), there are $2^n$ possibilities for $F_{1}$, as there are $2$ possibilities for each block in $F_{1}$. It is not required for an adversary to exactly reproduce the original $F_{1}$ for a given $F_{2}$. Any $F_{1}$ that maps to a given $F_{2}$ will compromise the system. Thus, even though the original template may be safeguarded, the process of biometric authentication is completely compromised as the system offers zero protection against an imposter trying to authenticate herself/himself to the system in contradiction to the aims of the authors  in \cite{WYang}. If an imposter tries to access the device, the authentication mechanism cannot detect it and thus cannot prevent intrusion.\\

\noindent Thus, in this case, we have shown that irrespective of the safety  of the  original biometric feature (that is, irrespective of whether the original biometric feature is hidden or not), nothing can be done to prevent an imposter from gaining access to the system without effort (there is no need to study the complexity of the best known algorithm to reverse the one-way function, as all that the imposter needs is the table above that can be constructed effortlessly).\\

\noindent It is straight forward to generalize the above attack to a BLOBLBS transformation of any segment size. In addition, the greater the blocksize, the lesser the chance of hiding the original biometric in the many-to-one transformation. For example, if the blocksize is as large as half of $F_{1}$, then there is a good $1$-in-$4$ chance of recovering the original template, if the chance of any bit in $F_{1}$ being a $0$ (or $1$) is $1$ in $2$.\\

\noindent (ii) As outlined above, cross matching between IoT devices results in a compromise of Privacy. Emphasizing this fact, the authors in \cite{WYang} write\\

\indent \textit{\dots What makes things worse is that other IoT devices may also use the same biometric template from the same user. It is well known that the original fingerprint feature data can never be changed, so the attacker could apply cross-matching on different IoT devices with the restored fingerprint information\dots}\\

\noindent It can be seen that the transformed template produced by the BLOBLBS algorithm is as vulnerable to cross-matching as the original biometric templates are. This is because, for a given input, the BLOBLBS algorithm will produce the same transformed template with every IoT device as the transformation function is the same irrespective of the device. In other words the transformed/protected templates across biometric databases can be linked and this compromises privacy (i.e., since protected templates across databases will be the same for a given entity, it could easily be inferred that it belongs to the same entity)\\

\noindent (iii)  The fundamental need of using the BLOBLBS algorithm is to enable canceling and re-establish protected biometric data of any system user. However, the BLOBLBS algorithm always produces the same transformed template. Thus a different new template cannot be generated and reissued when the previous one is to be retired. Thus the advantage of creating a transformed template in the first place no longer holds.\\

\noindent (iv) It is important that recognition accuracy is not impacted during the authentication process due to the information loss introduced by the protection mechanism. In fact, if $F_{1}$ is of size $1795$ bits and is sliced up into 5-tuples, since there are 359 such blocks, even without adversaries, many false positives can be generated, as $(2^{359}-1)$ combinations can produce as many matches. Recognition accuracy may also be impacted when the number of segments is computed as  \textit{floor(a/b)} where $a$ is the size of $F_{1}$ and $b$ is the segment size, especially when $b$ is large and when the filesize is not a multiple of the segment size. In this case, the number of segments can be taken to be \textit{roof(a/b)} and extra padding could be employed to fill in the gap in the last segment. Therefore, the absence of impact must be reported in big enough benchmarking datasets and with enough precision. The proposed system is only evaluated in FVC 2002~\cite{FVC2002}, and performance is reported only in terms of Equal Error Rate with a single significant digit, which is not enough to make a sensible comparison with the corresponding \textit{unprotected} system.\\

\noindent Other minor observations of \cite{WYang} include the following: 
The blocksize in the BLOBLBS algorithm is specified to be an odd number by the authors, but in experiments conducted, the authors specify the blocksize to be 3, 5 and 30 though 30 is not even an odd number!. 
\section{Conclusion}
\noindent In this article, we showed that it is possible to compromise the authentication process of a recently proposed Privacy Preserving Biometric Authentication System for IoT Security by authors in \cite{WYang}. In the process of doing this, we touched upon some aspects of security, for instance the ubiquitous one-way functions and the need for a rigorous description of a newly proposed security system. We also showed that individual privacy is compromised too when the recently proposed system is used.\\

\noindent While writing this article, we were reminded of what has now come to be known in the literature as \textit{Schneier's Law} which we paraphrase as follows:\\

\indent \textit{It is a safe idea to assume that a newly designed security system is unsafe unless there is evidence otherwise.}\\

\noindent Though the above law is intended to caution cryptography researchers and designers, the law may be applicable in other areas of security as well, such as biometrics. For more colorful versions of this law, we refer the reader to Bruce Schneier's blog \cite{Schneier}.

\end{document}